\documentclass[pra,twocolumn]{revtex4}

\usepackage[dvips]{graphicx}%
\usepackage{amsmath}
\usepackage{bm}

\begin{document}

\title{Cloning and optimal Gaussian individual attacks for continuous-variable quantum key distribution using coherent states and reverse reconciliation}%

\author{Ryo Namiki}
\email[Electric address: ]{namiki@qo.phys.gakushuin.ac.jp} 

\author{Masato Koashi}

\author{Nobuyuki Imoto}

\affiliation{CREST Research Team for Photonic Quantum Information, Division of Materials Physics, 
Department of Materials Engineering Science, 
Graduate school of Engineering Science, Osaka University, 
Toyonaka, Osaka 560-8531, Japan}

\date{\today}
\begin{abstract}
We investigate the security of continuous-variable quantum key distribution using coherent states and reverse reconciliation against Gaussian individual attacks based on an optimal Gaussian $1 \to 2$ cloning machine. We provide an implementation of the optimal Gaussian individual attack. We also find a Bell-measurement attack which works without delayed choice of measurements and has better performance than the cloning attack. 
\end{abstract}

\maketitle

\section{Introduction}

Quantum mechanical properties of physical systems make it possible to implement physically secure communication between distant parties, whereas it is impossible to achieve such a task only by the transmission of classical signals.  An interesting and actively investigated problem of the so-called quantum key distribution (QKD) protocols is how to achieve the secret-key sharing using imperfect semi-classical signals and devices \cite{rmp74}. One of the standard approaches is to implement photonic qubits by weak coherent states. QKD protocols have also been proposed based on quantum continuous-variable (CV) systems via coherent states \cite{coherent,coherentR,no-switch,hirano,postsel}. One of the central ideas in CV QKD is that the legitimate receiver (Bob) of the signal measures one of the conjugate quadratures randomly, although an exception is found \cite{no-switch}. There are two different types of CV-QKD protocols called postselection \cite{coherent,hirano} and reverse reconciliation (RR). In the RR protocol \cite{coherentR}, the sender (Alice) of the signal infers the measurement results of Bob to share the key.

In the classical picture, an amplifier (AMP) followed by a beam splitter (BS) provides perfect copies of the signal. Thus an eavesdropper (Eve) can obtain a perfect copy without making any disturbance on the signal. Then, by repeating this process, Eve can obtain arbitrary number of identical copies and characterize the signal with a desired resolution. In quantum theory, the amplification comes with spontaneous noise \cite{amp}. Thus the copies are imperfect and Eve's intervention can be detected due to the disturbance. In addition to this, her knowledge about the signal is limited by the imperfection in the copies. 

In CV systems, the AMP-followed-by-BS scheme provides an optimal Gaussian $1 \to 2$ cloning machine which makes the best approximate copies of Gaussian states \cite{optc1}. The impossibility of the cloning is sometimes connected with the security of QKD. Because a cloner makes a concrete example of eavesdropping attacks which always induces disturbance on the original system, one may think that the security against the best \textit{cloning attack} is a good measure of QKD performance. However, the conditions of the optimal cloning and optimal eavesdropping attack are in general different.
The optimal cloner provides two imperfect copies and one imperfect phase-conjugate-like state of the input called the \textit{anticlone}. 
In the cloning attack, Eve keeps one of the imperfect copies from which she intends to distill information, and discards the anticlone. It is natural to consider that Eve can perform a better attack by combining the clone-anticlone pair.

In connection with the clone-anticlone pair, there may exist several interesting quantum operations. For example, it is possible for Eve to erase the signal information and cancel the amplification noise by properly performing a Bell measurement on the clone-anticlone pair and displacing the other clone according to the measurement outcome as in CV quantum teleportation, although in this case Eve obtains no signal information \cite{reverse}. It is also known that quadrature signals can be efficiently encoded by using the phase conjugate pair of coherent states $| \alpha \rangle \otimes | \alpha ^* \rangle $ compared with the normal pair $| \alpha \rangle \otimes | \alpha  \rangle$ \cite{conju}. This suggests that Eve can also read the quadrature signal efficiently from the clone-anticlone pair. A variant of cloning machines that works with the phase-conjugate input has also been proposed \cite{pci}.

In this paper, we consider the security of CV QKD using coherent states and RR \cite{coherentR} against the cloning attack and its natural extensions, where, in addition to the clone, Eve also keeps and utilizes the anticlone. 
We show that an optimization of this attack corresponds to the optimal Gaussian individual attack \cite{coherentR,Gro03}. The realization of the optimal attack as well as that of the cloning attack seems to be impossible within the present technology because the attacks need a quantum memory to store the quantum signal for a sufficiently long time.
We find a Bell-measurement attack which works without a quantum memory and has better performance than the cloning attack.

\section{Eve's strategy}
\begin{figure}[htbp]
\includegraphics[width=1\linewidth]{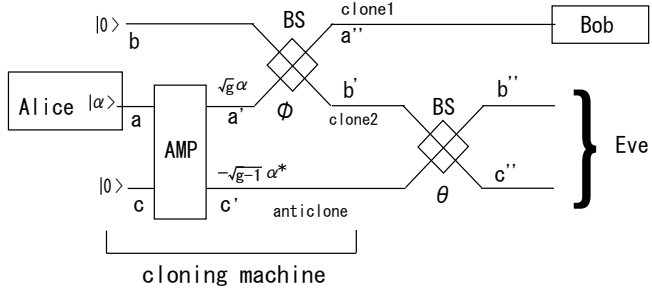}
\caption{Eve's attack is based on the Gaussian cloning machine which is constructed by a linear amplifier (AMP) followed by a beam splitter (BS). \label{fig1}}
\end{figure}

Let us consider a three-mode bosonic system and the quantum circuit including an AMP and two BSs as in Fig. \ref{fig1}. Alice prepares her state on the mode $  a$ which is the input mode of the cloner. Eve uses the ancillary modes $ b$ for the second clone and $ c$ for the AMP.

Eve's operation is as follows: 
First, Eve amplifies the input of the mode $a$ with the ancillary mode $c$.
Next, Eve combines the amplified mode $a'$ and the other ancillary mode $b$ by a BS with the transmission $\cos^2 \phi$. One of the outgoing clones (clone 1) on the mode $a''$ is received by Bob, and the other clone (clone 2) on the mode $b'$ is kept by Eve. After Eve learns Bob's measurement basis, she combines the anticlone on the mode $c'$, which comes from the ancillary mode of the AMP, and the clone 2 by a BS with the transmission $\cos^2 \theta$. Finally, Eve performs her measurement on $b''$ and $c''$.

 We define position quadrature and momentum quadrature of mode $k= \{  a,\   b, \   c, \ \cdots \}$ by 
\begin{eqnarray} 
\hat x _k  &\equiv& \frac{\hat k + \hat k ^\dagger }{2}, 
\ \hat p _k   \equiv \frac{\hat k - \hat k ^\dagger }{2i},
\end{eqnarray} where $\hat k $ ($\hat k ^\dagger$) is the annihilation (creation) operator of mode $k$. 

The transformations between modes are described by the unitary operators 
\begin{eqnarray} 
\hat   U_{  a,  c }
 (\lambda) &\equiv &e^{\lambda ( \hat a \hat c -\hat a ^\dagger \hat c^\dagger )}\end{eqnarray}
for an AMP with the amplification gain $g \equiv \cosh ^2 \lambda \ge 1$ and
\begin{eqnarray} 
\hat  V_{ a, b }
(\theta) &\equiv &e^{\theta ( \hat a \hat b^ \dagger  -\hat a ^\dagger \hat b )}
\end{eqnarray}  for a BS with the transmission $\cos^2
 \theta $.

The transformations between the modes are explicitly written as
\begin{eqnarray}
 \left ( \begin{array}{c}
 \hat a '    \\
 \hat c '
\end{array} \right) &\equiv& 
 \left ( \begin{array}{c}
    \hat U_{ a,  c }^{\dagger }(\lambda)  \hat a  \hat U_{ a, c }(\lambda)     \nonumber \\ 
 \hat U_{  a,  c }^{\dagger }(\lambda) \hat c   \hat U_{ a,  c }(\lambda) 
\end{array} \right) \\ \nonumber
&=& \left ( \begin{array}{c}
  \hat a \cosh \lambda -  \hat c ^\dagger  \sinh \lambda    \\
 \hat c  \cosh \lambda -  \hat a ^\dagger  \sinh \lambda   
\end{array} \right) \\ 
 &=&   \cosh \lambda  \left ( \begin{array}{cc}
 1 & 0 \\
  0 & 1 
\end{array} \right) \left ( \begin{array}{c}
  \hat a     \\
 \hat c    
\end{array} \right) - \sinh \lambda  \left ( \begin{array}{cc}
0 & 1 \\
  1 & 0 
\end{array} \right) \left ( \begin{array}{c}
  \hat a  ^\dagger     \\
 \hat c   ^\dagger      
\end{array} \right), \nonumber \\ \label{eq5}
\\ \nonumber
 \left ( \begin{array}{c}
 \hat a ''    \\
 \hat b '
\end{array} \right) &\equiv& 
 \left ( \begin{array}{c}
  \hat V_{ a',  b }^\dagger ( \phi )\hat a'\hat V_{ a', b } ( \phi )     \\ 
 \hat V_{ a',  b }^\dagger ( \phi ) \hat b  \hat V_{ a', b } ( \phi ) 
\end{array} \right) 
 \\ &=&    \left ( \begin{array}{cc}
 \cos \phi  & - \sin \phi \\
    \sin \phi &  \cos \phi 
  \end{array} \right) \left ( \begin{array}{c}
  \hat a'     \\
 \hat b    
\end{array} \right), \label{eq6}
\\\nonumber
 \left ( \begin{array}{c}
 \hat b ''    \\
 \hat c ''
\end{array} \right) &\equiv& 
 \left ( \begin{array}{c}
  \hat V_{ b' ,  c ' }^\dagger ( \theta )\hat b'\hat V_{ b ', c ' } ( \theta )     \\ 
 \hat V_{ b ', c ' }^\dagger ( \theta ) \hat c '  \hat V_{ b', c ' } ( \theta ) 
\end{array} \right) 
 \\ &=&    \left ( \begin{array}{cc}
 \cos \theta  & - \sin \theta \\
    \sin \theta &  \cos \theta  
  \end{array} \right) \left ( \begin{array}{c}
  \hat b '     \\
 \hat c '    
\end{array} \right).\label{eq7}
\end{eqnarray}
Combining Eqs. (\ref{eq5}),(\ref{eq6}) and (\ref{eq7}), we obtain the mode operators of the output:
\begin{widetext}
\begin{eqnarray} 
 \left ( \begin{array}{c}
 \hat a ''    \\
 \hat b ''   \\
 \hat c ''  
\end{array} \right) & = &\left ( \begin{array}{ccc} \cosh \lambda \cos \phi& - \sin\phi  &0 \\   \cosh \lambda \cos \theta  \sin \phi  & \cos \theta  \cos \phi & -\cosh \lambda \sin \theta \\   \cosh \lambda \sin \theta \sin \phi & \sin \theta \cos \phi  &\cosh \lambda \cos \theta   \end{array} \right)   \left ( \begin{array}{c}  \hat a       \\ \hat b       \\ \hat c       \end{array} \right) 
  - \sinh \lambda \left ( \begin{array}{ccc} 0 & 0   & \cos \phi \\  -\sin \theta & 0&  \cos \theta \sin \phi\\    \cos \theta & 0& \sin  \theta  \sin \phi \end{array} \right)   \left ( \begin{array}{c}  \hat a  ^\dagger     \\ \hat b   ^\dagger     \\ \hat c   ^\dagger       \end{array} \right). \label{maintrans}
\end{eqnarray}
\end{widetext}
Note that the state of Bob's mode $ a ''$ is determined by $\lambda$ and $\phi$, and does not depend on $\theta$.

In what follows, we use an abbreviated notation
\begin{eqnarray}
\langle    \hat F  \rangle_\alpha&\equiv & _c\langle 0 | _b\langle 0| _a\langle \alpha |  \hat F  |\alpha  \rangle_a |0 \rangle_b  |0 \rangle_c ,    \end{eqnarray}
where we defined the coherent state $| \alpha \rangle_k$ with the amplitude $\alpha$ of mode $k= \{a,b,c \}$ by $ \hat k | \alpha \rangle_k = \alpha | \alpha \rangle _k  $. 
For example, we can write the amplitudes of the output modes associated with the input $|\alpha \rangle_a $ as
 \begin{eqnarray}
 \langle   \hat a ''  \rangle_\alpha  &= & \alpha  \cosh \lambda \cos \phi ,  \\
\langle   \hat b ''  \rangle_\alpha &=&   \alpha  \cosh \lambda \cos  \theta \sin \phi + \alpha  ^* \sinh \lambda \sin \theta   ,\\
\langle   \hat c ''  \rangle_\alpha &=&   \alpha  \cosh \lambda \sin  \theta \sin \phi  - \alpha  ^* \sinh \lambda \cos \theta   \label{13}.
\end{eqnarray}
From these expressions, we can see that Eve's operation does not couple the real and imaginary parts of quadratures. 
This fact suggests that it is inefficient for Eve to measure a quadrature with an angle different from the one chosen by Bob. So we consider strategies in which Eve measures the same quadrature as Bob chooses.

We assume that Alice and Bob have the lossy and noisy transmission channel characterized by the line transmission $\eta$ and excess noise $\delta $, those are related to the amplitude and variance of Bob's mode $a''$ by  
\begin{eqnarray}
\langle  \hat a ''  \rangle_  \alpha &=& \sqrt \eta \alpha , \label{14}\\
\langle (\Delta x _{a''}) ^2 \rangle_  \alpha &=&\langle  \hat x _{a''} ^2 \rangle_  \alpha - \langle  \hat x _{a''}  \rangle_  \alpha^2  =
\frac{1}{4}(1 + \delta  ) . \label{15}
\end{eqnarray}
Suppose that Eve replaces the channel with the circuit in Fig. \ref{fig1}.  
Then, $\lambda$ and $\phi$ are determined to be
\begin{eqnarray}
\tan  \phi =\sqrt \frac{{1-\eta + \delta /2}}{{\eta - \delta /2}}, \label{phi}\\ 
\tanh \lambda = \sqrt \frac{\delta /2 }{\eta }.\label{lam}
\end{eqnarray}

If we take $\theta = 0$, $\eta =1 $, and $\delta =1$, the circuit is the optimal Gaussian $1 \to 2$ cloning machine \cite{optc1}. In this case we can see that the states of $a''$ and $b''$ are symmetric clones $\langle   \hat a ''  \rangle_\alpha = \langle   \hat b ''  \rangle_\alpha  =  \alpha$ and the clone-anticlone pair has the time-reversal relation $\langle   \hat c ''  \rangle_\alpha =   -\alpha^* =-\langle   \hat a ''  \rangle_\alpha^* $.

If we take $\theta = \pi /4 $, $\eta =1 $, and $\delta =1$, we can see that the mode operators of $b''$ and $c''$ almost duplicate the position and momentum quadrature of the input mode, respectively:
\begin{eqnarray}
\hat b ''&=&  \sqrt 2 \hat x_a + \frac{1}{2}\hat b - \hat c -\frac{1}{2}\hat c ^\dagger ,
\\
\hat c ''&=&   \sqrt 2 i \hat p_a + \frac{1}{2}\hat b + \hat c -\frac{1}{2}\hat c ^\dagger .
\end{eqnarray} This relation implies that by measuring $\hat x_{b''}$ and $\hat p_{c''}$ simultaneously Eve can efficiently learn the amplitude of the input state as in the efficient coding \cite{conju}.  
This measurement is considered to be a Bell measurement on the modes $b'$ and $c'$. As we will show later this Bell measurement leads to an efficient attack which does not need the quantum memory to store the signal coherently.

\section{ Reverse reconciliation protocol and conditional variances}
In the RR protocol, Alice sends the coherent state $| \alpha \rangle _a$ with the probability density
\begin{eqnarray}
P(\alpha )&=& \frac{2}{\pi V_A}e^{-\frac{2}{V_A}|\alpha |^2}
\end{eqnarray}
and Bob randomly measures one of the quadratures.
The density operator of the input state can be written as
\begin{eqnarray}
 \hat \rho &=& \int P(\alpha ) |\alpha \rangle_a|0\rangle_b |0\rangle_{cc} \langle 0|_b \langle0|_a \langle \alpha |d^2\alpha \label{dmt}.\end{eqnarray}
For calculation of the expectation values the following form is convenient
\begin{eqnarray}
\langle \hat F \rangle &\equiv& \textrm{Tr} \left( \hat F   \hat \rho  \right)
= \int P(\alpha )  \langle \hat F  \rangle_ \alpha d^2\alpha   .
\end{eqnarray} 

A sufficient condition for secure key distribution against a Gaussian individual attack where Eve uses the measurement result of a single mode quadrature after Bob's basis is declared is given by
\begin{eqnarray}  V(x_B| x_A)-  V(x_B| x_E)  & \le&  0, \nonumber \\ 
  V(p_B| p_A) - V(p_B| p_E)   &  \le&  0 , \label{cri2} \end{eqnarray}
where the conditional variance of $x$ given $y$ is defined by 
\begin{eqnarray}
V(x| y) &= & \langle  (\Delta  x )^2  \rangle -\frac{| \langle x y \rangle -  \langle x \rangle  \langle y \rangle|^2}{ \langle (\Delta y )^2  \rangle } \end{eqnarray} and the variables of Alice, Bob and Eve are denoted by the subscripts $A$, $B$ and $E$, respectively. The conditional variance becomes smaller as the variables become more correlated.

 Let us calculate Alice's conditional variances, $V(x_B| x_A) $ and $V(p_B| p_A)$. In our formulation, the first moment of the quadratures is $\langle \hat x_k \rangle =\langle \hat p_k \rangle =0$ for any $k$. The relevant terms for $V(x_B| x_A)$ can be written as 
\begin{eqnarray} 
\langle x_B ^2   \rangle &\equiv& \langle \hat x_{a''} ^2   \rangle =  \int P(\alpha )  \langle \hat x_{a''}^2 \rangle_ \alpha  d^2\alpha =\frac{1}{4}(1+\eta V_A +\delta ) 
\nonumber ,  \\ \label{eq3}\\
\langle x_A x_B   \rangle &\equiv& \int  P(\alpha ) \langle x_\alpha  \hat  x_{a''} \rangle_\alpha   d^2\alpha =
\frac{\sqrt \eta V_A}{4}  , \\
\langle x_A ^2   \rangle  & \equiv & \int     P(\alpha )  x_\alpha^2  d^2\alpha  = \frac{ V_A}{4} \label{e6}.
\end{eqnarray} 
Note that $x_\alpha \equiv \frac{\alpha + \alpha ^* }{2}$ corresponds to Alice's position quadrature which is determined by Alice's choice of the parameter $\alpha$. 
From Eqs. (\ref{eq3})-(\ref{e6}), we have
\begin{eqnarray}
V(x_B| x_A) =  \frac {1}{4} \left(1+\delta \right).  \label{vA}
\end{eqnarray} 
It means that Alice can predict Bob's measurement result within the uncertainty of the vacuum noise plus excess noise. This is also a direct consequence of Eq. (\ref{15}). Similarly we have
\begin{eqnarray}
V(p_B| p_A) =  \frac {1}{4} \left(1+\delta \right).  \label{vpA}
\end{eqnarray}

To calculate Eve's conditional variances, $V(x_B| x_E)$ and $V(p_B| p_E)$, let us assume that Bob measures $\hat x_{a''}$ and then Eve adjusts $\theta$ and measures $\hat x_{b''}(\theta )$. In this case, $ V(x_B| x_E) =  V(\hat x_{a''}|\hat x_{b''}(\theta))   $. From Eqs. (\ref{maintrans}), (\ref{phi}), and (\ref{lam}) we can write  \begin{eqnarray}
  \hat x_{a''} &=& \sqrt\eta \hat x _{a} - \sqrt{1- \eta +\delta /2} \hat x_{b} -\sqrt{\delta /2} \hat x_{c},\\
  \hat x_{b''} (\theta )&=& X \hat x _{a} + Y \hat x_{b} + Z \hat x_{c},
\end{eqnarray} 
where
\begin{eqnarray}
X &\equiv& \frac{1}{\sqrt{ \eta -\delta /2}}(\sqrt \eta  \sqrt{1- \eta +\delta /2}\cos \theta + \sqrt{ \delta /2} \sin \theta ) , \nonumber\\
Y &\equiv& \frac{1}{\sqrt{ \eta -\delta /2}}(  \eta-\delta /2 )\cos \theta \nonumber,  \\
Z &\equiv& -\frac{1}{\sqrt{ \eta -\delta /2}}(\sqrt \eta \sin \theta + \sqrt{ \delta /2} \sqrt{1- \eta +\delta /2}  \cos \theta )\nonumber  . \\  
\end{eqnarray} 
Then, using 
\begin{eqnarray}
\langle  \hat x_{b} ^2 \rangle &=&\langle  \hat x_{c} ^2 \rangle =\frac{1}{4} , \ 
\langle  \hat x_{a} ^2 \rangle = \frac{1}{4}(V_A +1),  \nonumber\\
\langle  \hat x_{a} \hat x_{b}  \rangle &=&\langle  \hat x_{b} \hat x_{c}  \rangle  = \langle  \hat x_{c} \hat x_{a}  \rangle =  0 , 
\end{eqnarray} we obtain
\begin{eqnarray}
\langle       x_{E} ^2  \rangle  &\equiv&  \langle      \hat x_{b''}(\theta ) ^2  \rangle  \nonumber \\ &=&    X^2 \langle   \hat x_{a }^2  \rangle + Y^2\langle   \hat x_{b }^2  \rangle +Z^2 \langle   \hat x_{c }^2  \rangle
 \nonumber \\ \label{e-pos}
&=& \frac{1}{4}\left(  X^2 (V_A +1) + Y^2 +Z^2
\right) ,\\
\langle       x_{E}  x_{B}   \rangle  &\equiv &  \langle     \hat x_{a''}  \hat x_{b''} (\theta )  \rangle  \nonumber \\
&=&  \sqrt \eta X  \langle     \hat x_{a}^2  \rangle  -Y \sqrt{1-\eta +\delta /2}   \langle   \hat x_{b }^2  \rangle   -Z \sqrt{ \delta /2}\langle   \hat x_{c }^2  \rangle   
\nonumber \\
&=&\frac{1}{4}\left(  \sqrt \eta X  (V_A +1) -Y \sqrt{1-\eta +\delta /2}   -Z \sqrt{ \delta /2}  
\right) .\nonumber \\ \label{eb-pos}
\end{eqnarray} 
From Eqs. (\ref{eq3}), (\ref{e-pos}) and (\ref{eb-pos}), we can write 
\begin{widetext}
\begin{eqnarray}
   V (\hat x_{a''}|\hat x_{b''}(\theta) )
 &= & V_{B|E}(\theta) 
  \equiv  \frac {1}{4} \left(1+\delta +\frac{\eta \left\{ V_A( X^2 +Y^2 +Z^2 -2X\Omega)-  \Omega^2  \right\} }{(V_A +1)X^2 + Y^2 +Z^2 }\right) \label{omegaa} 
\end{eqnarray}  with
\begin{eqnarray}\Omega &\equiv&  \left( X-  Y \sqrt{\frac{1-\eta +\delta /2}{\eta}}   -  Z \sqrt{ \frac{\delta }{2\eta}}  \right).  
\end{eqnarray}
\end{widetext}
Since the replacement of the modes $b'' \to c''$ corresponds to the replacement $\theta \to \theta - \pi /2 $ in Eq. (\ref{maintrans}), we can write
\begin{eqnarray}
  V (\hat x_{a''}|\hat x_{c''}(\theta) ) = V_{B|E}(\theta- \pi /2). \label{r1}
\end{eqnarray}

The conditional variance of momentum quadratures can be calculated in the same manner and we can verify  
\begin{eqnarray}
V (\hat p_{a''}|\hat p_{b''}(\theta) ) &=&V_{B|E}( - \theta) \label{vp1} , \\
V (\hat p_{a''}|\hat p_{c''}(\theta) ) &=& V_{B|E} (\pi /2 - \theta) . \end{eqnarray} 
Using relations (\ref{omegaa}), (\ref{r1}), (\ref{vp1}), and (\ref{vp2}), we can calculate $V(x_B|x_E)$ and $V(p_B|p_E)$ for various strategies taken by Eve.
From Eqs. (\ref{omegaa}) and (\ref{vp2}), we can see that by setting $\theta = \pi /4$ Eve can estimate both of the quadratures with the same uncertainty simultaneously, i.e., 
\begin{eqnarray}
V (\hat x_{a''}|\hat x_{b''}(\pi /4 ) ) &=&V (\hat p_{a''}|\hat p_{c''}(\pi /4 ) ) \nonumber  \\ \label{vp2} 
&=& V_{B|E} ( \pi /4 ).
\end{eqnarray}

\begin{widetext}
\section{cloning and Gaussian individual attacks}
In this section, we investigate the security condition (\ref{cri2}) for the following four attacks.
Firstly, we consider two simple attacks where Eve uses either the clone 2 or the anticlone. We call them the \textit{cloning attack} and the \textit{anticloning attack}, respectively.
Next, we consider two attacks where Eve uses both the clone 2 and the anticlone. In one attack, Eve does not use the quantum memory and performs a Bell measurement. 
In the other attack Eve optimizes $\theta$ in order to minimize her conditional variance.

\subsection{Cloning attack}
We assume that Eve utilizes only the mode $b'$ (clone 2).
In this case, $ V(x_B| x_E) =  V(\hat x_{a''}|\hat x_{b' } )   $ and $ V(p_B| p_E) =  V(\hat p_{a''}|\hat p_{b' } ) $. Using the relations $\hat x_{b'} = \hat x_{b''} (0) $, $\hat p_{b'} = \hat p_{b''} (0) $, (\ref{omegaa}), and (\ref{vp1}), we obtain
\begin{eqnarray}
V(x_B| x_E) = V(p_B| p_E)= V_{B|E} ( 0) &=& \frac{ \delta + 2 (1 + V_A) \eta  }{ \delta ^2 + \delta  \{ 1 +  (V_A -2) \eta \}   + 2 \eta \{ 1 + V_A(1 -\eta)\} } .
\end{eqnarray} 
The security condition (\ref{cri2}) requires
\begin{eqnarray}
\eta    \ge   \eta_{\textrm{clone}} &\equiv& \frac{\delta}{4V_A (1+\delta )}\left\{(3+ \delta ) V_A  -2 \delta+ \sqrt{\{(3+ \delta ) V_A  + 2 \delta \} ^2 + 16 V_A  }  \right\}  \nonumber \\
&= & \frac{\delta (3+ \delta )  }{4 (1+\delta )}\left\{1 - \frac{ \delta}{(3+ \delta ) V_A}+ \sqrt{ 1+  \frac{ 4 \delta   }{ (3+ \delta ) V_A\  }+  \frac{16V_A     + 4 \delta ^2}{  (3+ \delta )^2 V_A ^2 }    }  \right\} .\label{cri-i}
\end{eqnarray} 

\subsection{Anticloning attack}
We assume that Eve utilizes only the mode $c'$ (anticlone). In this case, $ V(x_B| x_E) =  V(\hat x_{a''}|\hat x_{c' } )   $ and $ V(p_B| p_E) =  V(\hat p_{a''}|\hat p_{c' } ) $. Using the relations $\hat x_{c'} = \hat x_{b''} (\pi /2) $, $\hat p_{c'} = \hat p_{b''} (\pi /2) $, (\ref{omegaa}), and (\ref{vp1}), we obtain
\begin{eqnarray}
V(x_B| x_E) = V(p_B| p_E)= V_{B|E} ( \pi /2) &=& 1+ \delta - \frac{ (4+3 V_A) \delta  - 2 V_A \eta  }{ (1+ V_A )\delta + 2\eta } . 
\end{eqnarray}
 We can write the security condition as
\begin{eqnarray}
\eta  &\ge &  \eta_{\textrm{anticlone}} \equiv  \frac{(4+3  V_A ) \delta }{2 V_A  } =  \left( \frac{ 2}{  V_A  }+ \frac{3}{2} \right)  \delta   \label{cri-ii} .
\end{eqnarray} 

\subsection{Bell measurement attack without delayed choice}
As we have seen in the ends of Sec. II and III, Eve may perform a Bell measurement on the clone-anticlone pair without using any quantum memory. Suppose that Eve's operation is as follows: Eve fixes $\theta = \pi /4$ so that the contributions of the mode $a$ to the modes $b''$ and $c''$ become equivalent. She performs position-quadrature measurement on the mode $b''$ and momentum-quadrature measurement on the mode $c''$ right after she received the state. She chooses one of the measurement results after she learns Bob's choice of the quadratures. We call this attack the \textit{Bell measurement attack} (BMA). In this case, $ V(x_B| x_E) =  V(\hat x_{a''}|\hat x_{b'' } (\pi /4) )   $ and $ V(p_B| p_E) =  V(\hat p_{a''}|\hat p_{c'' } (\pi /4) ) $. Thanks to the relation (\ref{vp2}), we can estimate Eve's conditional variances by
\begin{eqnarray}
V_{B|E} ( \pi /4) &=& 1+ \frac{  V_A (2 \eta - \delta  )^2  }{ \delta ^2 +2 (V_A +2) \sqrt{2\eta \delta (1 -\eta + \delta /2) } +  \delta  \{ V_A +2 +  (V_A -2) \eta \}   + 2 \eta \{2 + V_A(1 -\eta)\} }.  \nonumber \\
\end{eqnarray} 
The security condition becomes
\begin{eqnarray}
\eta   \ge    \eta_{\textrm{BMA}} &\equiv & \frac{2\delta \left(V_A +1 -\delta /2 + \sqrt{ (V_A +2 )(V_A -\delta )} \right) }{V_A (2+ \delta ) } \nonumber \\
&= &\frac{2\delta}{2+\delta } \left( 1+ \frac{1-\delta /2}{V_A} +\sqrt {1+\frac{2 -\delta }{V_A} - \frac{2\delta}{ V_A^2 } }  \right)  \label{cri-iii}.
\end{eqnarray} 

\end{widetext}

\subsection{Optimal Gaussian individual attack} 
In our formulation, Eve's optimal strategy is to select $\theta $ in order to minimize her conditional variance. The minimum value of $V_{B|E}$ is given by
\begin{eqnarray}
V_{B|E} (\theta_{\textrm{opt}} ) &=& \frac{1+ V_A}{(1+V_A)(1+ \delta)- \eta V_A } 
\end{eqnarray}with
\begin{eqnarray}
\theta_{\textrm{opt}} &\equiv& \tan ^{-1}\frac{\sqrt{ \eta  \delta } (2+V_A)}{\sqrt{2-2 \eta +\delta } \{V_A(\eta -\delta ) -\delta \} } , \nonumber \\ && (-\pi /2 < \theta_{\textrm{opt}} < \pi /2 ). 
\end{eqnarray} 
Eve can achieve the optimal value for each quadrature by measuring $\hat x _{b''}(\theta_{\textrm{opt}} )$ or $\hat p _{b''}(-\theta_{\textrm{opt}} )$ according to Bob's choice, position or momentum quadrature, respectively;
\begin{eqnarray}   V(x_B| x_E) &=&  V(\hat x_{a''}|\hat x_{b'' }(\theta_{\textrm{opt}}  ) )=V_{B|E}(\theta_{\textrm{opt}}  )    \nonumber \\
 V(p_B| p_E) &=&  V(\hat p_{a''}|\hat p_{b'' } (- \theta_{\textrm{opt}} ) ) =V_{B|E}(\theta_{\textrm{opt}}  ).     
\end{eqnarray} 

The value $V_{B|E} (\theta_{\textrm{opt}}  )$ corresponds to the lower bound of Eve's conditional variance which saturates the Heisenberg-type uncertainty relation \cite{coherentR,Gro03,comment} and thus the optimization of our scheme gives an implementation of the optimal Gaussian individual attack. Another implementation of the optimal Gaussian individual attack is found in \cite{Gro03}. 

The security condition can be written as
\begin{eqnarray}
 \eta  \ge \eta_{\textrm{opt}} \equiv \frac{1+ V_A}{V_A}  \frac{ \delta (2+ \delta)}{(1+ \delta  ) }  .
 \label{cri-iiii}
\end{eqnarray}

In the high-modulation limit $(V_A \to \infty )$, we have
\begin{eqnarray}
V_{B|E} (\theta_{\textrm{opt}} ) &\to & \frac{1}{1+\delta-\eta} , \nonumber \\
\theta_{\textrm{opt}}   & \to & \tan ^{-1}\frac{\sqrt{ \eta  \delta }}{\sqrt{2-2 \eta +\delta } (\eta -\delta )  }.
\end{eqnarray} 

\subsection{Discussion}

\begin{figure}[htbp]
\includegraphics[width=1\linewidth]{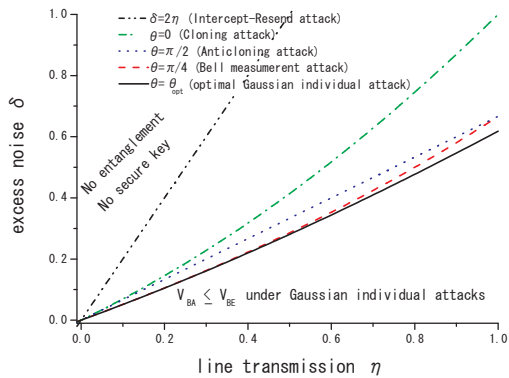}
\caption{The security condition is shown for the line transmission $\eta $ and quadrature excess noise $\delta $ in the high-modulation limit ($V_A \to \infty$). The dash-dotted line is for the cloning attack, the dotted line is for the anticloning attack, the dashed line is for the Bell-measurement attack, and the solid line is for the optimal Gaussian individual attack. The dot-dash-dotted line is for the intercept-resend attack which gives a necessary condition of CV QKD using coherent states \cite{namiki2,namiki3}. 
  \label{Fig-d-e} }
\end{figure}

In Fig. \ref{Fig-d-e}, we show the curves given by $\eta = \eta_{\textrm{opt}}$, $\eta = \eta_{\textrm{BMA}}$, $\eta = \eta_{\textrm{anticlone}}$, and $\eta = \eta_{\textrm{clone}}$ at the high-modulation limit $(V_A \to \infty )$. For each attack, the security condition is satisfied below the curve. We also show a necessary condition of CV QKD using coherent states \cite{namiki2,namiki3}: $ \eta > \delta /2 $. This bound is given by an intercept-resend attack based on the simultaneous measurement of the quadratures and can be also derived from the separable condition of CV systems \cite{Gro03,comment}. Above this curve secure key distribution is impossible. 

From Fig. \ref{Fig-d-e} we can see that the anticloning attack is better than the cloning attack and BMA is better than the anticloning attack at the high-modulation limit. This order is conserved in the case of finite modulation provided $V_A \ge \delta $ and $\delta \le 2/3 $, i.e., from Eqs. (\ref{cri-i}), (\ref{cri-ii}), (\ref{cri-iii}) and (\ref{cri-iiii}) we can verify  
\begin{eqnarray}
 \eta_{\textrm{opt}} \ge \eta_{\textrm{BMA}}
 \ge \eta_{\textrm{anticlone}} \ge  \eta_{\textrm{clone}} \label{fin}. 
\end{eqnarray} 
 It shows that the cloning attack is the weakest attack among the four. The fact that the cloning attack is weaker than the anticloning attack seems to be counter-intuitive because the amplitude of the clone 2 is always larger than that of the anticlone if $ \delta \le 2 \eta $, i.e.,
\begin{eqnarray}
\left| \frac{\langle   \hat b '  \rangle_\alpha }{\langle   \hat c '  \rangle_\alpha } \right| \ge 1.
\end{eqnarray} An interpretation of the result is as follows: 
One can show that the clone 1 and clone 2 are not entangled but in a mixture of coherent states \cite{namiki3}. Thus the bipartite system cannot make stronger correlations between any two of the quadratures beyond the vacuum fluctuation. On the other hand, since the amplified mode and anticlone mode, $a'$ and $c'$, are entangled due to the process of the AMP, it is possible to make stronger correlations between the quadrature of them beyond the vacuum fluctuation. This leads to a smaller conditional variance.

Relation (\ref{fin}) also shows that the clone-anticlone pair without delayed choice of measurements provides a better attack than either of the cloning and anticloning attacks. This implies the utility of the phase-conjugate pair \cite{reverse,conju,pci}.  

In the high-modulation and high-loss limit ($V_A \gg 1 $ and, $\eta  \ll 1 $), from Eqs. (\ref{cri-i}), (\ref{cri-ii}), (\ref{cri-iii}), and (\ref{cri-iiii}) we can see that the cloning and anticloning attacks provide nearly the same security bound
\begin{eqnarray}
 \eta_{ clone} \sim \eta_{\textrm{anticlone}} \sim  \frac{3}{2}   \delta ,
 \end{eqnarray} and that BMA provides nearly optimal bound: 
\begin{eqnarray}
 \eta_{\textrm{BMA}} \sim  \eta_{\textrm{opt}} \sim 2\delta .  
\end{eqnarray} 
The existence of such an effective attack without delayed choice of measurements is interesting and it seems to be a characteristic of CV QKD using coherent states.
In general, indirect measurement is considered to be powerful because Eve can use a quantum memory and she can perform her measurement after she learns Bob's measurement basis. If Eve cannot use the quantum memory she has to measure different quantities simultaneously to infer the signal because she does not know the basis. Generally the simultaneous measurement of non-commutable observables costs additional noise and indirect measurement without delayed choice seems to be inefficient. However, in our case of the CV QKD under realistic condition, the difference between with and without the quantum memory is not significant as shown above. From practical aspect BMA should be considered seriously because it means that a very efficient attack can be realized without advanced technologies.

\section{Summary}
We have investigated the security of CV QKD using coherent states and reverse reconciliation against individual Gaussian attacks based on an optimal Gaussian 1$\to$2 cloning machine. We have assumed that one of the clones is delivered to Bob and Eve combines the other clone and the anticlone using a BS and performs quadrature measurements. 
In this approach, we can connect and analyze different individual Gaussian attacks just by changing the parameter of the BS.

We found an efficient individual attack which works without delayed choice of measurements and has better performance than either of the cloning and anticloning attacks. The operation of this attack provides a simultaneous-measurement scheme which leaves an imperfect copy. In other words, it provides a way to perform a simultaneous measurement of the quadratures indirectly so that the measurement induced noise corresponds to a given value. In the high-loss and high-modulation limit, we showed that the security bound given by this attack is nearly the same as the one given by the optimal Gaussian individual attack. We also found that the optimization of the BS parameter with a delayed choice of measurements provides an implementation of the optimal Gaussian individual attack.



\begin{thebibliography}{}
\bibitem{comment} In Ref \cite{coherentR,Gro03}, the excess noise is defined by $  \epsilon =  \delta /\eta  $.
\end{thebibliography}
\end{document}